\documentclass[english,prd,twocolumn,superscriptaddress,nofootinbib,preprintnumbers]{revtex4}
\usepackage[latin1]{inputenc}
\usepackage{graphicx}
\usepackage{amssymb}
\usepackage{color}
\usepackage{float}
\usepackage{amsmath}
\usepackage{amsfonts}
\usepackage{dcolumn}
\usepackage{hyperref}

\newcommand{\rd}{{\rm d}}

\begin{document}

\title{Light mass galileons: Cosmological dynamics, mass screening and observational constraints}

\author{Amna Ali}
\affiliation{Astroparticle Physics and Cosmology Division and Centre
for Astroparticle Physics, Saha Institute of Nuclear Physics, 1/AF
Bidhannagar, Kolkata 700 064, India}

\author{Radouane Gannouji}
\affiliation{Department of Physics, Faculty of Science, Tokyo
University of Science, 1-3, Kagurazaka, Shinjuku-ku, Tokyo 162-8601,
Japan}

\author{Md. Wali Hossain}
\affiliation{Centre for Theoretical Physics, Jamia Millia Islamia, New Delhi-110025,
India}

\author{M. Sami}
\affiliation{Kobayashi-Maskawa Institute for the Origin of Particles
and the Universe, \\Nagoya University, Nagoya 464-8602, Japan}
\affiliation{Centre for Theoretical Physics, Jamia Millia Islamia, New Delhi-110025,
India}

\begin{abstract}

In this paper, we examine the cosmological viability of
 a light mass galileon field consistent with local gravity constraints. The minimal, $L_3=\Box\phi(\partial_\mu \phi)^2$,
  massless
 galileon field requires an additional term  in order to give rise
 to a viable ghost free late time acceleration of Universe. The
 desired
  cosmological dynamics can either be achieved by incorporating an additional terms in the action such
  as $(L_4,L_5)$ $-$ the higher order galileon Lagrangians
  or by considering a light mass field {\it \`a la} galileon field potential. We analyse the second
  possibility and find that: (1) The model produces a viable
  cosmology in the regime where the non-linear galileon field is
  subdominant, (2) The Vainshtein mechanism operates at small scales where the non-linear effects become
  important and contribution of the field potential ceases to be significant.
  Also the small mass of the field under consideration is protected against strong quantum
corrections thereby providing quantum stability to the system.
\end{abstract}

\maketitle

\section{Introduction}
Modified theories of gravity pose a serious alternative to dark
energy, an exotic cosmic fluid, needed to account for late time
cosmic acceleration in the framework of standard lore\cite{review1,
vpaddy,review2,review3,review3C,review3d,review4,review5}. The task
of building an alternative to Einstein's theory is very challenging
as the latter fits with the observation with a great accuracy
locally thereby a large scale modification is either felt locally or
the framework gets reduced to $\Lambda CDM$. A viable alternative
theory of gravity should satisfy important requirements: (1) It
should be close to $\Lambda CDM$ but yet distinguishable from it,
(2) Theory should be free from ghost and tachyon instabilities and
(3) Theory should not conflict with local physics. The third
requirement is often very stringent and its compliance needs the
invoking of special mechanisms. The modified theories necessarily
include additional scalar degree(s) of freedom in some form or the
other. And to be relevant to late time cosmic acceleration, it
should be light mass entity which, on the other hand, could cause a
havoc as a fifth force never seen in the laboratory  or in the solar
neighborhood. It therefore necessary to screen out the effect of the
fifth force in a delicate manner.

Broadly, there are two methods of mass screening, the chameleon
mechanism and the Vainshtein effect. The chameleon scenario
\cite{Khoury:2003aq} relies on the direct coupling of  matter to
scalar field with potential. The mass of the field becomes dependent
on the density of environment which justifies the designation
"chameleon" for such a field. The chameleon potential is chosen such
that the effective mass of the field increases with local matter
density thereby leading to suppression of fifth force. The chameleon
mechanism  is an extremely powerful tool for mass screening but as
noticed by many authors (see e.g. \cite{Babichev:2009fi}) the
numerical integration of the system for a spherically
 symmetric background hits singularity because
 of the form of the chameleon potential. Hence an extreme
 fine-tuning of the initial conditions is required to avoid the problem.
 The chameleon theories are also plagued with the problem of large
 quantum corrections because of the large mass of the chameleon field required
 to pass the local gravity tests \cite{Upadhye:2012vh}.

 The Vainshtein mechanism \cite{Vainshtein:1972sx} is a superior field theoretic method
 of mass screening. It was invented by Vainshtein in 1972
 to address the discontinuity problem in massive gravity of
 Pauli-Fierz. It relies on non-linear derivative term of the type
 $L_3=(\partial_\mu \phi)^2\Box \phi$ where $\phi$ is scalar degree of
 freedom of helicity zero graviton in this case. The
 dynamics of non-linear term gives rise to a miraculous
 phenomenon: Around a massive body, in a large radius dubbed
 Vainshtein  radius, fifth force is suppressed switching off any
 modification to gravity locally. The field is strongly coupled to itself and hence becomes weakly coupled to matter sector. The DGP model \cite{Dvali:2000hr} contains such
 a non-linear term in the so called decoupling limit responsible
 for the compliance of the model with local gravity constraints \cite{Luty:2003vm}.
 The non-standard kinetic term also occurs in Kaluza-Klein reduction
 of Gauss-Bonnet gravity to four dimensional space time \cite{Gannouji:2011qz} which
 makes clear why the theory is free from Ostrogradsky ghosts. The role
 of the scalar degree of freedom is played by dilaton in this case.
 The field $\phi$ due to the presence of an underlying symmetry in flat space time
 was termed as galileon. There exist higher order galileon
 Lagrangians $L_4$ and $L_5$ that contain higher order non linear
 derivative
 terms in which four and five galileon fields participate
 respectively. Recently, more general galileon action were
 constructed \cite{deRham:2010eu,Goon:2011qf} and their cosmological
 implications were investigated \cite{DeFelice:2010pv}. They belong to a more general class of models first introduced in
\cite{Fairlie:1991qe}.

The galileon field with the lower order term $L_3$ is sufficient to
take care of the local gravity constraints whereas $L_5$ does not
contribute to mass screening and though $L_4$ can effect the
numerics of Vainstein radius but adds nothing to the underlying
physics of Vainshtein mechanism.

However, galileon system with $L_3$ term alone can not give rise to
late time acceleration of universe. It was first demonstrated in
\cite{Gannouji:2010au} that at least one higher order
Lagrangian, say $L_4$ \cite{Nicolis:2008in} be added to action in order to produce a
stable de Sitter solution. This is the analogue of the DGP model, where the
self-accelerating branch is unstable because of the presence of ghost \cite{Koyama:2005tx}. Also
physical implications of the cubic Galileon term coupled non-minimally to
the metric was first studied in \cite{Chow:2009fm,Silva:2009km}. Since we
are working in a
phenomenological setting, we could restrict ourselves to $L_3$ but
add a potential to galileon to produce late time cosmic
acceleration. We do not assign a field theoretic mechanism to
produce the potential for galileon, perhaps one would need some
non-perturbative machinery to do the job.

Our proposal can be seen as an attempt to replace chameleon
mechanism by the Vainshtein effect: In chameleon scenario, the
choice of specific form of chameleon potential is crucial to the
model which needs to be extremely fine-tuned locally for the
successful implementation of the underlying chameleon ideology.

\section{ Basic setup and equations of motion}
It is well known that in General theory of Relativity, the Newtonian
and longitudinal gravitational potentials are equal provided that
anisotropic stress tensor is absent. Modified gravity theories
usually predict a difference between these two potentials $-$ a
gravitational slip. At the linear level of perturbations, the
modified theories of gravity give rise to an anisotropic stress
tensor which is  really not a signature of a particular modified
theory as usually claimed. At the linear level of perturbations, the
modified theories of gravity give rise to an anisotropic stress
tensor which is  really not a distinctive signature of  modification
of gravity as usually claimed. A modified theory of gravity
can, in general, be described in terms of an imperfect  fluid giving
rise to a gravitational slip. Thus we can not distinguish whether
the slip is caused by the presence of an imperfect fluid or by the
modification of gravity. Since the observations, at present, do not
show any difference between the gravitational potentials (see e.g.
\cite{Daniel:2009kr}), it is interesting to study models that do not
produce a gravitational slip.

Following the results derived in \cite{DeFelice:2011hq}, we deduce
that the most general action with second order differential
equation for the metric and the scalar field which do not produce a slip is of one of the following form

\begin{align}
\mathcal{S}_1&=\int d^4x \sqrt{-g} \Bigl[F(\phi) R+\Bigl(F'(\phi)\ln X+G(\phi)\Bigr)\Box\phi\nonumber\\
&+H(\phi)\Bigl((\Box\phi)^2-\phi_{,\mu\nu}\phi^{,\mu\nu}+XR-\frac{3}{2} \nabla_{\mu}\Bigl(\phi^{,\mu}\Box\phi\Bigr)\nonumber\\
&- G_{\mu\nu}\phi^{,\mu}\phi^{,\nu}\Bigr)+K(\phi,X)\Bigr]
\end{align}
or
\begin{align}
&\mathcal{S}_2=\int d^4x \sqrt{-g} \Bigl[R+G(\phi,X)\Box\phi+K(\phi,X)\nonumber\\
&+F(\phi)\Bigl((\Box\phi)^2-\phi_{,\mu\nu}\phi^{,\mu\nu}+XR- G_{\mu\nu}\phi^{,\mu}\phi^{,\nu}\Bigr)\Bigr]
\end{align}

where $X=-\frac{1}{2}g^{\mu\nu}\phi_{,\mu}\phi_{,\nu}$.

The two actions are different but in both cases we do not have a gravitational slip.
The second action $S_2$ is a generalization of the KGB model studied in
\cite{Deffayet:2010qz}. We  also note that for $S_1$ in the
Brans-Dicke form $\phi R$, the presence of $\ln X~\Box\phi$ is
important to cancel the slip between the two potentials.

Keeping in mind a  well motivated model which appears in the
decoupling limit of DGP, we shall consider the second action
hereafter.
 We also introduce a coupling of field to matter, therefore we work in the
Einstein frame where the theory does not gives rise to a slip.
Obviously in the Jordan frame the same theory would produce a slip
because of the conformal factor as noticed in \cite{Brax:2010tj}.
The conformal transformation would produce terms which are not in
the form of the original action.

The second action is the most general action free from Ostrogradsky
ghost problem, it does not produce a gravitational slip and contains
the decoupling limit of DGP as a sub-class. But the first
action can be seen as a modified Brans-Dicke model which can produce
interesting local and cosmological solutions.

In what follows, we shall consider a simple model with $G\equiv X$
which corresponds to the decoupling limit of DGP, $F\equiv 0$ and $K\equiv
X-V(\phi)$. With these choices, our action has the form (in Einstein
frame)
\begin{align}
S=\int d^4x\sqrt{-g}\Bigl [\frac{M^2_{\rm{pl}}}{2} R &-  \frac{1}{2}(\nabla \phi)^2\Bigl(1+\frac{\alpha}{M^3}\Box \phi\Bigr) - V(\phi) \Bigr]\nonumber\\
&+ \mathcal{S}_m\Bigl[\psi_m;e^{2 \beta \phi/M_{pl}} g_{\mu\nu}\Bigr]
\label{1.1}
\end{align}
where $M_{\rm{pl}}^2=8\pi G$ is the reduced Planck mass, $M$ is a
energy scale, $(\alpha,\beta)$ are dimensionless constants and $V$
is the potential for
the field. \\
This action corresponds to the coupled quintessence field with a
galileon like correction $(\nabla \phi)^2 \Box \phi$. Variation of
the action (\ref{1.1}) gives the following equations of motion
\begin{align}
M^2_{\rm{pl}} G_{\mu\nu}= T_{\mu\nu}^{(m)}+T_{\mu\nu}^{(r)}+T_{\mu\nu}^{(\phi)}\label{1.2}
\end{align}
\begin{align}
& \Box \phi+\frac{\alpha}{M^3}\Bigl[(\Box\phi)^2-\phi_{;\mu\nu}\phi^{;\mu\nu}-R^{\mu\nu}\phi_{;\mu}\phi_{;\nu}\Bigr]\nonumber\\
&-V'(\phi)=-\frac{\beta}{M_{\rm{pl}}}T^{(m)}
\label{1.3}
\end{align}

where
\begin{align}
T_{\mu\nu}^{(\phi)}&=
\phi_{;\mu}\phi_{;\nu}-\frac{1}{2}g_{\mu\nu}(\nabla\phi)^2
-g_{\mu\nu}V(\phi)\nonumber\\
&+\frac{\alpha}{M^3} \Bigl[\phi_{,\mu}\phi_{;\nu}\Box\phi+g_{\mu\nu}\phi_{;\lambda}\phi^{;\lambda\rho}\phi_{;\rho}-\phi_{;(\mu}\phi_{;\nu)
\rho}\phi^{;\rho}\Bigr]
\end{align}

where $'$ denotes the derivative wrt $\phi$.

In the discussion to follow, we shall investigate the cosmological
dynamics  and local gravity constraints in the model based upon the
equations of motion (\ref{1.3}).


\section{Fifth Force, Screening effect and Local Physics}
In our model, in general, we might have both the screening
mechanisms operating due to the presence of potential and the
galileon term. We wish to arrange things in such a way that
potential contributes negligibly to local physics giving way to
Vainshtein mechanism. The potential should come into action at large
scales giving rise to cosmic acceleration.

We now consider spherically symmetric, static solution of
equation (\ref{1.3}). In flat Minkowski background and in presence of
non relativistic matter, we have
\begin{align}
\frac{1}{r^2}\frac{\rd}{\rd r}\left[r^2\phi'(r)\right]+\frac{2\alpha} {M^3r^2}\frac{\rd}{\rd r}\left[r\phi'(r)^2\right]-\frac{\rd V{(\phi)}}{\rd \phi}=
\frac{\beta}{M_{pl}} \rho_m
\label{fifth}
\end{align}

We consider that the same phenomenon is at the origin of late time
acceleration and local suppression of fifth force in the action. We therefore associate to both the non-linear term and the
potential the same energy scale $M$. We have $\phi \simeq M$ and
$V\simeq M^4$. As we  previously mentioned, the chameleon mechanism
generates huge fine-tuning conditions. Thus if we want to prefer the
Vainshtein mechanism, we should demand that at small scales
$\frac{2\alpha} {M^3r^2}\frac{\rd}{\rd r}\left[r\phi'(r)^2\right]
\gg \frac{\rd V{(\phi)}}{\rd \phi}$. Then considering $\alpha \simeq
1$ we may have a negligible potential contribution for scales $L \ll
1/M$. This scale should  at least be larger that the apogee which
tells us that $M\ll 10^{-15}$eV. Also we should assume
that the potential term is negligible compared to the matter part.
Considering $\rho_m \simeq 1$g$/$cm$^{-3}$, we have $M\ll
10^{-3}$eV. This condition could  always be verified as soon as the
first one is fixed. We shall see later that for viable cosmology,
the potential should be of the order of critical density which
implies $M\simeq 10^{-3}$eV. It seems
 contradictory with the previous statement. In fact, from the energy scale $M$, the results will
 also depend on the form of
 the potential.
 We found that for inverse power law potentials, we
 are in the  chameleon regime and
 hence divergences can occur. We will take up this issue in the next section where we shall consider the
 class of potentials in detail which give  negligible
  contribution locally for $M\simeq 10^{-3}$eV.

In that case,  neglecting the potential, the system reduces to the
standard  Galileon with  $L_3$. At large scales, the potential is
dominant and this part would be studied in the next section. We note
there is also an intermediate regime where the potential is of the
order of $L_3$. We would not consider the analysis of this regime
here. In fact at very large distances, the approximations made in
this section can not be trusted, the expansion of the Universe
should be taken into account too. Last but not least, as there is no
physical system isolated at large scales,  it is not known how the
screening mechanism would operate for $N$ particles  if each one of
them possess its own Vainshtein radius $-$ the so-called {\it
elephant} problem.

For the scales of interest mentioned above, the fifth force is

\begin{align}
F_{\phi}=\frac{\beta M^3 r}{4\alpha M_{pl}}\left[-1+\sqrt{1+\Bigl(\frac{r_V}{r}\Bigr)^3}\right]
\label{analytical}
\end{align}

where $r_V$ is the Vainshtein radius of  the  body with mass $M$
and $r_s$ is the Schwarzschild radius.

\begin{align}
 r_v=\left(\frac{8\alpha\beta M_{pl}r_s}{M^3}\right)^{\frac{1}{3}}
\label{vradius}
\end{align}

The Vainshtein radius is defined as the distance from where onwards
the fifth force becomes comparable with the gravitational force.
Considering the Sun, we need a Vainshtein radius larger than
the size of solar system which implies that
$M<10^{-7}(\alpha\beta)^{1/3}$ eV. The scale $M$ plays a
fundamental role in local physics determined by Vainshtein
mechanism. This is also  the scale associated with the potential
thereby  to the cosmology of the model. In order to get late time
acceleration of the Universe, we need to fix, $M\simeq 10^{-3}$ eV.
On the other hand,  in order to have Vainshtein radius larger than the solar system, we should demand that $\alpha\beta>10^{12}$.
 Another option would be to choose $\alpha\simeq 1$ or $M^4 \ll \Lambda$ where $\Lambda$
 is the cosmological constant. This would make the model even less natural than the cosmological
 constant in terms of an unnatural value. Perhaps,  it  points towards a possibility that the scale
 could be
 linked to some large extra dimension \textit{\`a la} DGP model. Finally, one could also
  consider two different scales $(M_{1},M_{2})$ in the model associated with the galileon term and
 the potential respectively. We shall not consider  this possibility here.

In our analysis we have considered two different potentials

\begin{align}
V(\phi)=M^4\left[e^{-\frac{\mu_1 \phi}{M_{pl}}}+e^{-\frac{\mu_2\phi}{M_{pl}}}\right]
\label{pot1}
\end{align}
and
\begin{align}
V(\phi)=M^4\left[\cosh\left(\frac{\gamma
\phi}{M_{pl}}\right)-1\right]^m, \label{pot2}
\end{align}

where $\lambda$, $\mu$, $\gamma$ are constants and $m$ is a positive
integer. In both cases, we found numerically that the potential is
negligible locally and that
the eq.(\ref{fifth}) can be approximated by a vanishing potential.\\
The system appears as the decoupling limit of DGP and as it is well
known that the Vainshtein mechanisms protects the model in a static
spherically symmetric configuration. In the next section we shall
demonstrate  that the potential is dominant at large scales and
gives rise to a correct cosmology contrary to the same model without
potential \cite{Gannouji:2010au}.

\section{On the bound from the time variation of $G$}

All the theories described in the Einstein frame by a
 time-varying coupling to matter leads to a variation
 of the effective Newton constant ($G$) as seen in the
  Jordan frame. There are two effects which modify $G$
  \cite{Babichev:2011iz}:
  The exchange of helicty-0 modes which is suppressed in our
   model because of the Vainshtein mechanism and the rescaling of
   the coordinates because of the conformal transformation. In our case the time
    variation of $G$ occurs because of the rescaling factor only. Hence it is easy to see that
\begin{align}
|\dot G/G| \approx \beta\dot{\phi}/M_{pl}.
\end{align}
It was shown in Ref.\cite{Babichev:2011iz} that a model involving
only derivatives of the fields would give
\begin{align}
|\dot G/G|_\text{today}\approx \beta H_0.
\end{align}
But the observational constraints from the Lunar Laser Ranging (LLR)
gives $|\dot G/G|_\text{today}<0.01 H_0$ \cite{Williams:2004qba} which translates into a
restriction on $\beta$, namely, $\beta<0.01$. In the presence of a
potential, the situation is drastically modified. In fact, in this
case, we can assume that the field sits at the minimum of the
effective potential. As a result,  in case of the potential
(\ref{pot1}), we have
\begin{align}
|\dot G/G|_\text{today}\approx \frac{6\beta^2}{\beta^2+\mu_1^2+\mu_2^2} H_0
\end{align}
Hence the LLR constraint gives
\begin{align}
\beta \lesssim \frac{\sqrt{\mu_1^2+\mu_2^2}}{25}
\end{align}
In the next section, we will consider  cosmology for
$(\mu_1=20\gg\mu_2)$ which means that $\beta\leq 1$. Therefore we
see that the LLR bound will not bring additional constraints as soon
as the standard cosmological constraints are satisfied.


\section{Cosmological dynamics of the model}

In a spatially flat  FLRW background, the equations of motion take
the form

\begin{align}
3M_{\rm{pl}}^2H^2 &=\rho_m+\rho_r+\frac{\dot{\phi}^2}{2}\Bigl(1-\frac{6\alpha}{M^3} H\dot{\phi}\Bigr)+V{(\phi)}\,,\\
M_{\rm{pl}}^2(2\dot H + 3H^2)&=-\frac{\rho_r}{3}-\frac{\dot{\phi}^2}{2}\Bigl(1+\frac{2\alpha}{M^3}\ddot{\phi}\Bigr)+V(\phi)\,,\\
-\frac{\beta}{M_{\rm{pl}}} \rho_m &=\ddot{\phi}+3H\dot{\phi}-\frac{3\alpha}{M^3} \dot{\phi}\Bigl(3H^2\dot{\phi}+\dot{H}\dot{\phi}+2H\ddot{\phi}\Bigr)\nonumber\\
&+V'(\phi).
\end{align}

The equation of conservation which can be derived from the previous equations is
\begin{align}
\dot\rho_m+3H\rho_m &=\frac{\beta}{M_{\rm{pl}}}\dot{\phi} \rho_m ,\\
\dot\rho_r+4H\rho_r &=0
\end{align}


Let us introduce the following dimensionless quantities

\begin{align}
x&=\frac{\dot{\phi}}{\sqrt{6}H M_{\rm{pl}}}\,,\quad y=\frac{\sqrt{V}}{\sqrt{3} H M_{\rm{pl}}}
\label{xy} \\
\epsilon &=-6\frac{\alpha}{M^3}H\dot \phi\,, \quad \lambda=-M_{\rm{pl}}\frac{V'}{V}
\label{epsilon}
\end{align}

needed to cast the evolution equations in the form of an autonomous
system

\begin{align}
\frac{{\rm d}x}{{\rm d}N}&=x\Bigl(\frac{\ddot{\phi}}{H\dot{\phi}}-\frac{\dot H}{H^2}\Bigr)\\
\frac{{\rm d}y}{{\rm d}N}&=-y \Bigl(\sqrt{\frac{3}{2}}\lambda x+\frac{\dot H}{H^2}\Bigr)\\
\frac{{\rm d}\epsilon}{{\rm d}N}&=\epsilon \Bigl(\frac{\ddot{\phi}}{H\dot{\phi}}+\frac{\dot H}{H^2}\Bigr)\\
\frac{{\rm d}\Omega_r}{{\rm d}N}&=-2\Omega_r\Bigl(2+\frac{\dot H}{H^2}\Bigr)\\
\frac{{\rm d}\lambda}{{\rm d}N}&=\sqrt{6}x\lambda^2(1-\Gamma)
\end{align}

where $N\equiv \ln a$, $\Gamma=\frac{VV_{,\phi\phi}}{V_{,\phi}^2}$ and

\begin{align}
\frac{\dot H}{H^2}&=\frac{2(1+\epsilon)(-3+3y^2-\Omega_r)-3x^2(2+4\epsilon+\epsilon^2)}{4+4\epsilon+x^2\epsilon^2}\nonumber\\
&\qquad \qquad \qquad \qquad +\frac{\sqrt{6}x\epsilon (y^2\lambda -\beta \Omega_m)}{4+4\epsilon+x^2\epsilon^2}\\
\frac{\ddot{\phi}}{H\dot{\phi}}&=\frac{3x^3\epsilon-x\Bigl(12+\epsilon (3+3y^2-\Omega_r)\Bigr)+2\sqrt{6}(y^2\lambda-\beta\Omega_m)}{x(4+4\epsilon+x^2\epsilon^2)}
\end{align}

In the case of an exponential form of the potential, $\Gamma=1$ and $\lambda$ is a constant.
Therefore the system reduces to the set of four equations as usual.

For $\epsilon=0$, we recover the standard coupled dark-energy model
\cite{Amendola:1999er}. The system has the same dynamical phase
plane as the coupled quintessence except one additional de-Sitter
solution, for $(\epsilon=-2,\lambda=0,y^2-x^2=1)$. This solution
exists only if the equation $\lambda^2\Gamma(\lambda)=0$ admits the
solution $\lambda=0$. Before getting to numerical analysis,
we should broadly summarize the constraints on the model parameters.
we have parameters $(\mu_1,\mu_2)$ in the potential (\ref{pot1})
which control the different phases of the dynamics. One of the
parameters $\mu_1$, gives rise to the radiation era whereas, $\mu_2$
is responsible for the late time cosmic acceleration.
 The matter era is principally controlled by $\beta$.
 In order to have viable thermal history $(\Omega_r=1-4/\mu_1^2)$,  we need to impose a constraint on
 $\mu_1$. Indeed, the nucleosynthesis  demands that $\Omega_\phi=4/\mu_1^2\leq 0.01$ during the radiation
 era  which imposes restriction on $\mu_1$ , namely, $\mu_1\geq 20$. For stable acceleration
 of the Universe we need to impose constraints on other parameters
 such that,
  $\mu_2<\sqrt{2}$ and also $\beta<(3-\mu_2^2)/\mu_2$. During the Dark Energy
  phase,
  $\omega_{\text{eff}}=-1+\mu_2^2/3$. We shall make use of this information while carrying out the
  numerical analysis. Indeed,  in the  numerics to follow,
  we will take small values for $\beta$ and adhere to $(\mu_1,\mu_2)=(20,0.5)$. During the
  evolution, $\lambda$ would change from $\mu_1$ to $\mu_2$ hence if $\mu_2$ is
  small enough, it would imply a small variation of the potential
  which would give rise to quasi-de-Sitter universe, $\omega_{\text{eff}}\simeq -1$.

\begin{figure}[h]
\centering
\includegraphics[scale=.6]{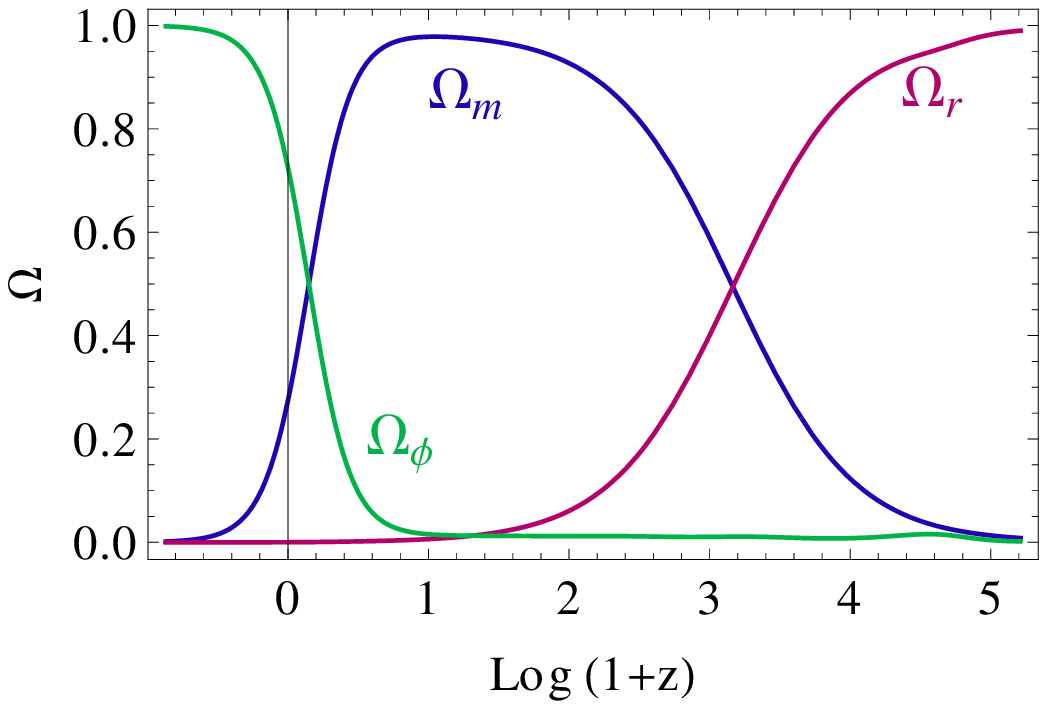}
\includegraphics[scale=.6]{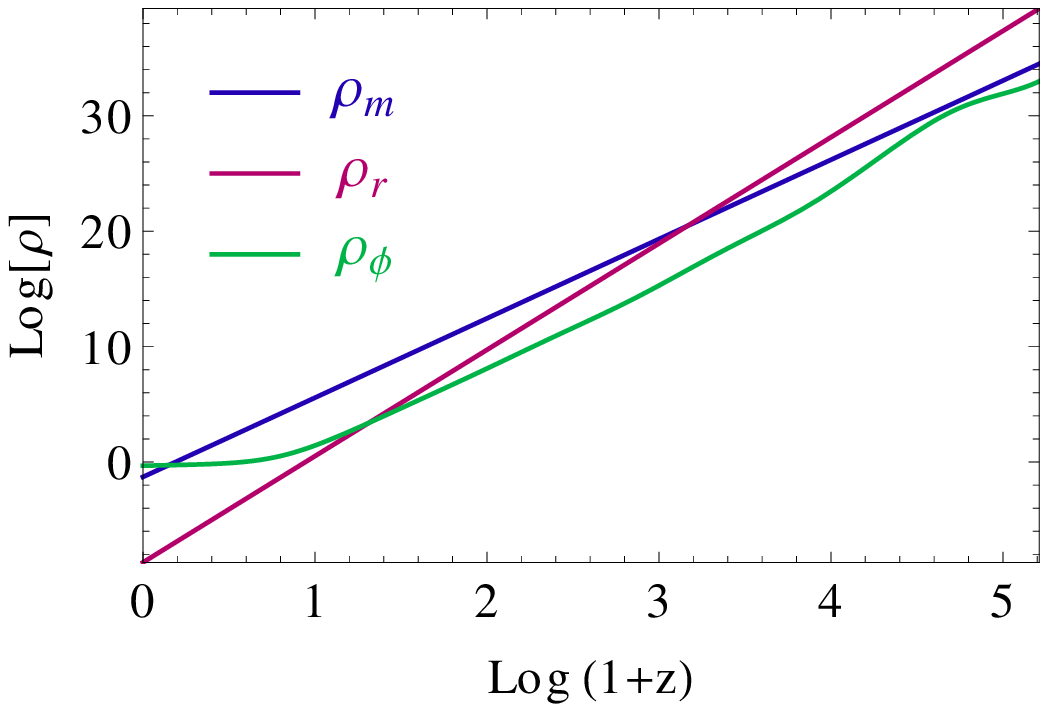}
\caption{(Top): Density parameters $\Omega$ of matter, radiation and galileon field for potential (\ref{pot1}) are shown here with $\mu_1=20$ , $\mu_2=0.5$ and $\beta=0.1$. (Bottom): Density $\log\Bigl(\frac{\rho}{3M_{pl}^2H_0^2}\Bigr)$ for the 3 fields and the same parameters $(\mu_1,\mu_2,\beta)$.}
\label{Fig1}
\end{figure}

\begin{figure}[h]
\centering
\includegraphics[scale=0.6]{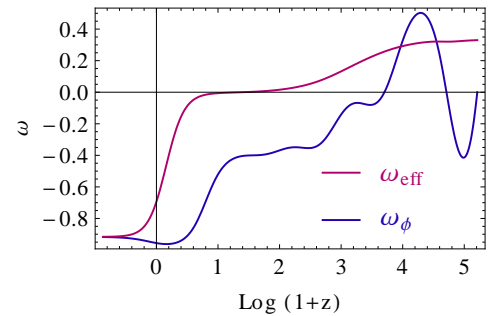}
\caption{The cosmic evolution of the field equation of state $w_\phi$ and the total effective equation of state $w_\text{eff}$ for $\mu_1=20$ , $\mu_2=0.5$ and $\beta=0.1$.}
\label{Fig2}
\end{figure}

We have numerically analysed the autonomous system and find that the
cosmology of the model is similar to the case of coupled
quintessence Fig.(\ref{Fig1},\ref{Fig2}). The influence of the
higher derivative term disappears fast with evolution leaving behind
the coupled quintessence. We have not listed the various fixed
points and their corresponding eigenvalues to avoid the repetition
of the earlier work on the similar theme. We have plotted the
dimensionless density parameter and field energy density versus the
scale factor on the $\log$ scale for the double exponential
potential. The plots show in case of the overshoot a tracker
behaviour. The parameters can be conveniently fixed to produce a
viable late time cosmology.


\section{Cosmological Perturbation}

In this section we analyse the perturbations of the model. As
mentioned in the introduction, the action is built in such away that
the two gravitational potentials are same. Hence we consider the
following metric

\begin{equation}
\ ds^2=-(1+2\Psi)dt^2+a(t)^2(1-2\Psi)d\vec{x}^2
\label{pmetric}
\end{equation}

In the subhorizon approximation, we have

\begin{equation}
 \ddot{\delta}_m+(2H+\frac{\beta}{M_{pl}}\dot{\phi})\dot{\delta}_m-4\pi G_{eff}\rho_m\delta_m=0
\label{growth}
\end{equation}
Where,
\begin{equation}
 G_{eff}=G\left(1+\frac{\left(\frac{\alpha}{M^3M_{pl}}\dot{\phi}^2+2\beta\right)^2}{2-\frac{\alpha^2}{M^6M_{pl}^2}\dot{\phi}^4-8\frac{\alpha}{M^3}H\dot{\phi}-
4\frac{\alpha}{M^3}\ddot{\phi}}\right)
\label{Geffth}
\end{equation}

where the gauge-invariant density contrast is defined as $\delta_m=\frac{\delta \rho_m}{\rho_m}+3Hv$ and $v$ is the velocity of the fluid.

\begin{figure}[h]
\includegraphics[scale=.6]{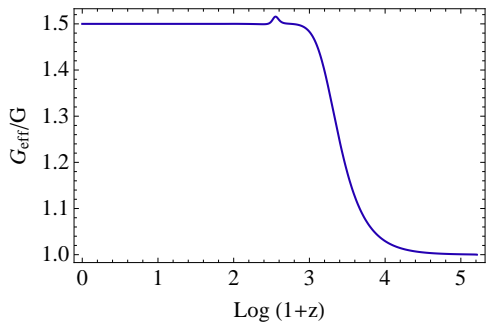}
\caption{Evolution of $G_\text{eff}/G$ for the potential (\ref{pot1}) with $\mu_1=20$, $\mu_2=0.5$ and $\beta=0.5$.}
\label{Geff}
\end{figure}

In Fig.(\ref{Geff}), we show the evolution of $G_\text{eff}$
as a function of
 the redshift. Because of the non-linear term
 of the galileon field, we have at large redshift
 $G_\text{eff}$ similar to General Relativity which
 is consistent with BBN constraints, whereas $G_\text{eff}=G(1+2\beta^2)>G$ in
 the coupled quintessence model. Thus the non-linear term of the galileon
 field can alleviate constraints on the model.
We have seen that the bound from the LLR is weak, hence we
should expect
 to have large values of the coupling  $\beta$.
 However,  we shall notice in the section to follow that the model under consideration reduces to
 coupled quintessence scenario at large scales thereby giving rise to strong cosmological
 constraints on $\beta$ as first noticed  in Ref.\cite{Amendola:1999er}.

\section{Data analysis}

To get some insight into the parameter window for the models, we constrain them
using Supernovae, BAO and CMB data. We have also used the data of the growth factor $f$ \cite{growth} defined as

\begin{equation}
 f=\frac{{\rm d} \ln \delta_m}{{\rm d} \ln a}
 \label{grth}
\end{equation}

We show the evolution of the growth factor $f$ for the two
  potentials studied in this paper. In both cases,
  the deviation from $\Lambda$CDM model is not significant.
  We should also mention that $f>1$ at $z=3$ which means that
    the growth is faster than in Einstein-de-Sitter model. This is
    opposite to the prediction of Dark Energy scenario. This sounds very
   strange but since the deviation from $\Lambda$CDM is not
   significant, the latter does not effect the statistical analysis.
    Only future observations can conform this result.

\begin{figure}[h]
\includegraphics[scale=.6]{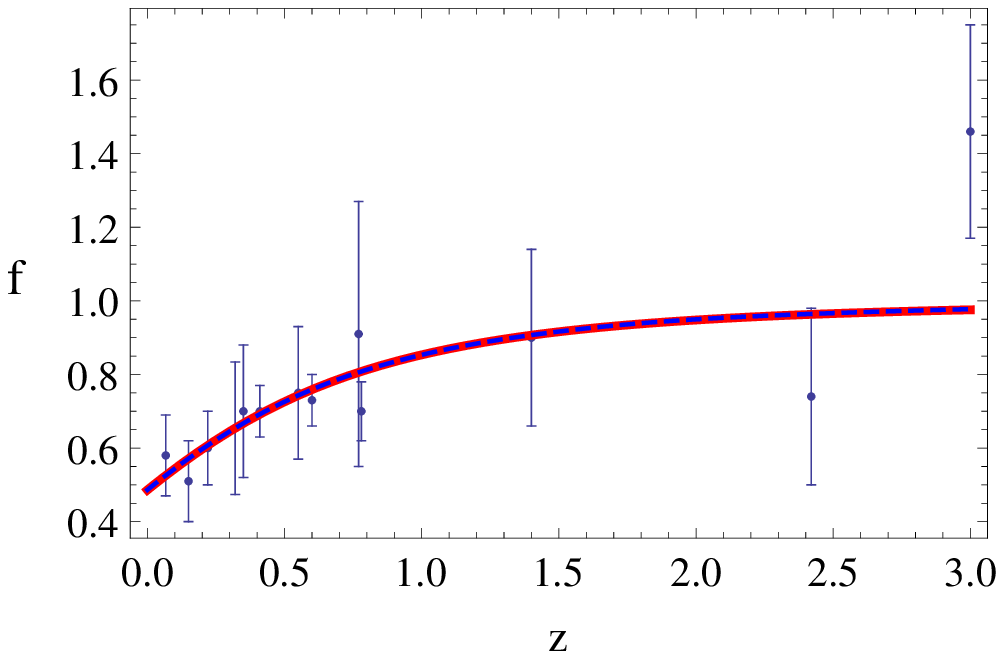}
\includegraphics[scale=.6]{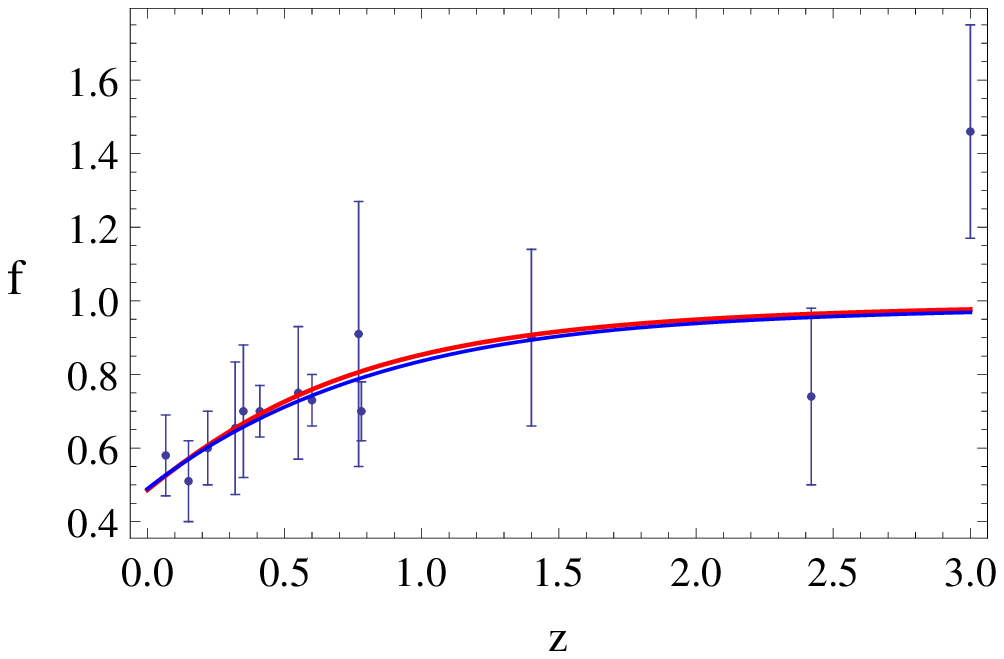}
\caption{Evolution of the growth factor $f$ for potential (\ref{pot1}) and (\ref{pot2}) respectively.
The integration is done from $\Omega_m\approx 0.98$ for both the cases with $\beta=0.1$. For the potential (\ref{pot1}) we have taken $\mu_1=20$ and $\mu_2=0.5$ and for the potential (\ref{pot2}) we have taken $\gamma=30$ and $m=0.5$. The red line corresponds to $\Lambda$CDM model.}
\label{growth}
\end{figure}

We define
\begin{align}
\chi^2=\chi_{Growth}^2+\chi_{SN}^2+\chi_{BAO}^2+\chi_{CMB}^2.
\end{align}

where $\chi_{Growth}^2$ is defined as,

\begin{align}
\chi_{Growth}^2(\theta)=\sum_i \frac{f_{obs}(z_i)-f_{th}(zi,\theta)}{\sigma_f(z_i)}
\end{align}

Where, $f_{\text{obs}}$ is the observational data and $\sigma_f$ is the 1$\sigma$ error. $f_{th}$ is the theoretically calculated value of growth factor from the eqn$(\ref{grth})$. $\theta$ is the model parameter.

For the SN type Ia observation we consider Union2.1 data compilation \cite{Suzuki:Union2.1} of 580 data points. The observable for the SN type Ia is the
distance modulus $\mu$ which is defined as, $\mu=m - M=5 \log D_L+\mu_0$, where $m$ and $M$ are the apparent and absolute magnitudes of the
Supernovae and luminosity distance $D_L(z)=(1+z) \int_0^z\frac{H_0dz'}{H(z')}$ and $\mu_0=5 \log\left(\frac{H_0^{-1}}{M_{pc}}\right)+2
5$ is a nuisance parameter and should be marginalized. For this observation the $\chi^2$ is defined as,

\begin{align}
\chi_{SN}^2(\mu_0,\theta)=\sum_{i=1}^{580} \frac{\left(\mu_{th}(z_i,\mu_0,\theta)-\mu_{obs}(z_i)\right)^2}{\sigma_\mu(z_i)^2}
\end{align}

Where, $\sigma_{\mu}$ is the uncertainty in the distance modulus. After marginalizing $\mu_0$ as \cite{Lazkoz:2005sp} we can have,

\begin{align}
\chi_{SN}^2(\theta)=A-\frac{B^2}{C}
\end{align}

Where,

\begin{align}
&A(\theta) =\sum_{i=1}^{580} \frac{\left(\mu_{th}(z_i,\mu_0,\theta)-\mu_{obs}(z_i)\right)^2}{\sigma_\mu(z_i)^2}\\
&B(\theta) =\sum_{i=1}^{580} \frac{\mu_{th}(z_i,\mu_0,\theta)-\mu_{obs}(z_i)}{\sigma_\mu(z_i)^2}\\
&C(\theta) =\sum_{i=1}^{580} \frac{1}{\sigma_\mu(z_i)^2}
\end{align}

\noindent For BAO we have used the data of $\frac{d_A(z_\star)}{D_V(Z_{BAO})}$ \cite{BAO}, where $z_\star$ is the decoupling time $z_\star \approx 1091$, the
comoving angular-diameter distance $d_A(z)=\int_0^z \frac{dz'}{H(z')}$ and $D_V(z)=\left(d_A(z)^2\frac{z}{H(z)}\right)^{\frac{1}{3}}$. Required data for
this analysis is given in table $\ref{baodata}$.

\begin{center}
\begin{table*}
\begin{tabular}{|c||c|c|c|c|c|c|}
\hline
 $z_{BAO}$  & 0.106  & 0.2 & 0.35 & 0.44 & 0.6 & 0.73\\
\hline \hline
 $\frac{d_A(z_\star)}{D_V(Z_{BAO})}$ &  $30.95 \pm 1.46$ & $17.55 \pm 0.60$ & $10.11 \pm 0.37$ & $8.44 \pm 0.67$ & $6.69 \pm 0.33$ & $5.45 \pm 0.31$  \\
\hline
\end{tabular}
\caption{Values of $\frac{d_A(z_\star)}{D_V(Z_{BAO})}$ for different values of $z_{BAO}$.}
\label{baodata}
\end{table*}
\end{center}

\noindent $\chi_{BAO}^2$ is defined as,

\begin{equation}
 \chi_{BAO}^2=X_{BAO}^T C_{BAO}^{-1} X_{BAO}
\end{equation}
Where,
\begin{equation}
X_{BAO}=\left( \begin{array}{c}
        \frac{d_A(z_\star)}{D_V(0.106)} - 30.95 \\
        \frac{d_A(z_\star)}{D_V(0.2)} - 17.55 \\
        \frac{d_A(z_\star)}{D_V(0.35)} - 10.11 \\
        \frac{d_A(z_\star)}{D_V(0.44)} - 8.44 \\
        \frac{d_A(z_\star)}{D_V(0.6)} - 6.69 \\
        \frac{d_A(z_\star)}{D_V(0.73)} - 5.45
        \end{array} \right)
\end{equation}

and the inverse covariance matrix,

\begin{widetext}
\begin{align}
C^{-1}=\left(
\begin{array}{cccccc}
 0.48435 & -0.101383 & -0.164945 & -0.0305703 & -0.097874 & -0.106738 \\
 -0.101383 & 3.2882 & -2.45497 & -0.0787898 & -0.252254 & -0.2751 \\
 -0.164945 & -2.45499 & 9.55916 & -0.128187 & -0.410404 & -0.447574 \\
 -0.0305703 & -0.0787898 & -0.128187 & 2.78728 & -2.75632 & 1.16437 \\
 -0.097874 & -0.252254 & -0.410404 & -2.75632 & 14.9245 & -7.32441 \\
 -0.106738 & -0.2751 & -0.447574 & 1.16437 & -7.32441 & 14.5022
\end{array}
\right).
\end{align}
\end{widetext}

Finally we used the CMB shift parameter $R=H_0 \sqrt{\Omega_{m0}} \int_0^{1089}\frac{dz'}{H(z')}$.
For this $\chi_{CMB}^2$ is defined as,

\begin{align}
 \chi_{CMB}^2(\theta)=\frac{(R(\theta)-R_0)^2}{\sigma^2}
\end{align}

Where, $R_0=1.725 \pm 0.018$ \cite{CMB}.

\begin{figure}[h]
\centering
\includegraphics[scale=.6]{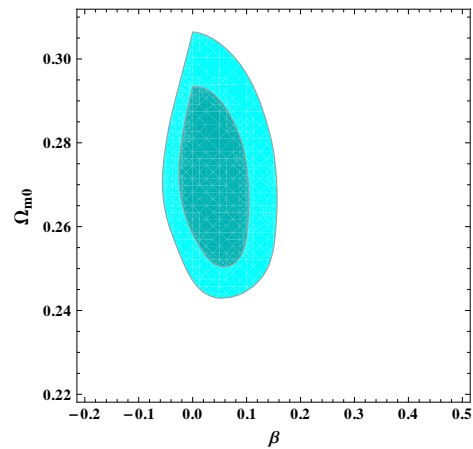}
\caption{The 1$\sigma$ (inside) and 2$\sigma$ (outside) likelihood contours in the $(\Omega_{m,0},\beta)$ plane for total $\chi_{Growth+SN+BAO+CMB}^2$.}
\label{chitot}
\end{figure}

\begin{figure}[h]
\centering
\includegraphics[scale=.7]{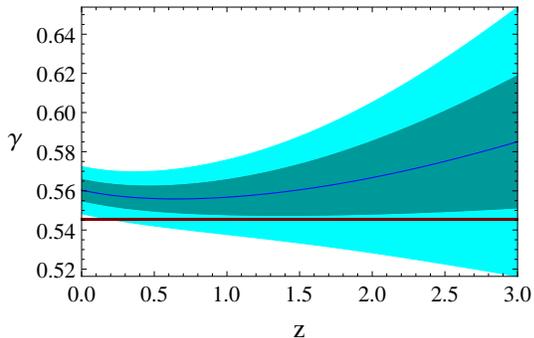}
\caption{The 1$\sigma$ (inside) and 2$\sigma$ (outside) of the growth index. The central line is the growth index with the best fit values of the model parameters $\beta$ and $\Omega_{m0}$ in the case of the double exponential (\ref{pot1}).}
\label{growthindex}
\end{figure}

In this analysis we have two model parameters $\beta$ and $\Omega_{m0}$. We varied $\beta$ from $-0.5$ to $1$ and $\Omega_{m0}$ from $0.2$ to $0.33$. Figure (\ref{chitot}) shows the results in the $(\beta,\Omega_{m0})$ parameter space. The analysis is performed for $M=10^{-3}$ eV. Minimum value of the total $\chi^2$ is at $\beta \sim 0.04$ and $\Omega_{m0} \sim 0.275$. Using these best fit values of the model parameters growth index $\gamma$ is plotted in the figure (\ref{growthindex}) with the $1\sigma$ and $2\sigma$ errors. At late time, the evolution is consistent with Dark Energy models where $w>-1$ \cite{Polarski:2007rr}. But the model shows a larger value of $\gamma$ for all redshift compared to $\Lambda$CDM. This is a different characteristic than $f(R)$-gravity models \cite{Gannouji:2008wt} or scalar-tensor theories \cite{Gannouji:2008jr} where $\gamma<0.55$.

We find that $\beta \simeq 0$, $\beta$ large is not a problem for local physics (chameleon, Vainshtein)
 but at large scales we need to consider $\beta$ small as in coupled quintessence.
 In fact, the galileon term is negligible at large scales and we recover the standard
 coupled quintessence model.

\section{Conclusion}
In this paper, we have investigated a particular case of a
more general class of models which do not produce a gravitational
slip. This model includes  the $L_3$-term derived in the
decoupling limit of DGP and a galileon field potential added to the
Lagrangian on phenomenological grounds. The potential breaks the
galilean symmetry in a flat spacetime but serves as an important
toll  at large scales for obtaining  a viable cosmology. The scale
in the model and the parameters are carefully set such that the
galileon term is dominant at small scales which suppress the effect
of the chameleon effect leaving the Vainshtein mechanism to operate.
This was one of the motives of our proposal to replace the chameleon
mechanism as the latter is plagued by the problems of fine tuning
and large quantum corrections
 due to  the large mass of the chameleon field. In contrast, the domination
  of the galileon term at small scales successfully generates a screening effect without requiring
  a large mass of the field and the quantum corrections remain small. We have, therefore shown that models
  including light mass scalar fields coupled to matter become viable as soon as we introduce
  a lower order galileon, term, $(\nabla \phi)^2\Box\phi$ in the action.
  We should
  also emphasize that
  this term does not significantly modify the background cosmology of the model.
  Our model is  built in away as not to produce large deviation
  from standard model for cosmological perturbations due to the
  vanishing of the gravitational slip. This approach can be generalized
  to a large class of light mass scalar field models which are otherwise ruled out
  by local gravitational constraints.\\
We have demonstrated that the model passes the local tests and can
give rise to viable cosmology at late times.  It produces a small
deviation of the growth factor compared to
 $\Lambda$CDM model. It interesting to note that in case of pure
 coupled quintessence, $G_{eff}=G(1+2\beta^2)$ thereby requiring  coupling
 to be small in order to respect  the BBN constraint. On the other
 hand, in the model under consideration, $G_\text{eff}=G$ as
 in General Relativity which certainly alleviates the constraints on coupling at large redshifts.
 In principle, our scenario involves
 cosmological constraints at low redshift which
 depends on the form of the potential as usual.

\section{ACKNOWLEDGEMENTS}
The work of RG is supported by the Grant-in-Aid for Scientific Research Fund of the JSPS No. 10329.
 MWH acknowledges the funding from CSIR, Govt of India. MWH wants to thank Sumit Kumar for useful discussions on statistical techniques. MS thanks S. Nojiri and K. Bamba for useful discussions.
  MS is supported by JSPS
fellowship and thanks Kobayashi-Maskawa Institute for the Origin of
Particles and the Universe for hospitality. MS is also supported by
the Department of Science and Technology, India under project No.
SR/S2/HEP-002/2008.

\end{document}